\documentclass[12pt]{article}
\usepackage{amsfonts} 
\def\be{\begin{eqnarray}}
\def\ee{\end{eqnarray}}
\def\*{\star} 
\begin{document}
\phantom{.}

\centerline{\large A CLASSICAL BOUND ON QUANTUM ENTROPY} 

\phantom{.}

\phantom{.}

\centerline{ \Large Cosmas K Zachos }
$^{\sharp}$ High Energy Physics Division,
Argonne National Laboratory, Argonne, IL 60439-4815, USA \\
\phantom{.} \qquad\qquad{\sl zachos@hep.anl.gov}      

\begin{abstract}
A classical upper bound for quantum entropy is identified and illustrated, 
$0\leq S_q \leq \ln ( {e \sigma^2  / ~ 2\hbar} )$,
involving the variance $\sigma^2$ in phase space of the 
classical limit distribution of a given system. A fortiori, this 
further bounds the corresponding information-theoretical generalizations of 
the quantum entropy proposed by R\'{e}nyi.  
\end {abstract}
\vskip 0.5cm

\hrule

\vskip 0.5cm

\section{Introduction}

Recurrent problems in four dimensional BPS black holes focus on the 
entropic behavior of  the respective complex structure moduli spaces, and, 
perhaps independently, on the corresponding holographic entanglement 
information lost in decoherence, 
and associated Hawking radiation paradoxes \cite{ooguri}. 
They all rely on the fundamental and dependable statistical concept
of entropy, which accounts collectively for the flow of information in 
these systems, and for which robust estimates are needed, in lieu of detailed 
accounts of quantum states. Ideally, such estimates would only require 
gross geometrical and semiclassical
features of the system involved, and ignore quantum mechanical interference 
subtleties. 

Classical continuous distributions have been studied in probability
and information theory for quite a long time, and Shannon \cite{ash} 
has derived handy upper bounds for their entropy, and thus crude least 
information estimates, in the 1940s. 
Approximate counting of quantum microstates, however, is normally toilsome, 
and can be approximated heuristically by semiclassical proposals \cite{wehrl}, 
which, ultimately, should devolve to a bona fide classical limit, 
despite occasional ambiguities and complications along the way \cite{takabayasi}. 
However, a more systematic approach was initiated 
by Braunss \cite{braunss},
who appreciated the 
underlying simplicity of phase space in 
taking a classical limit of intricate quantum systems. 
He thus tracked the information loss involved in smearing away quantum effects, 
to argue that the entropy of a quantum system is majorized by 
that of its classical limit, 
as $\hbar$-information of the former is forfeited in the latter, an intuitively 
plausible relation. 

The purpose of the present brief remark is to simply combine the two 
inequalities into a general upper bound of the quantum entropy of a system 
provided essentially by just the logarithm of the variance in phase space of 
the classical limit distribution of that system.
The resulting inequality, eqn (\ref{new}) below, 
is illustrated simply by the elementary physics paradigm of a 
thermal bath of oscillator excitations of one degree of freedom, 
whose phase-space representation is an obvious maximal entropy Gaussian. 

Note that there is no specific assumption of a particular spectral 
behavior---or even of the existence of a hamiltonian---for the systems 
covered by the inequality. 
Extension to arbitrary degrees of freedom and tighter bounds contingent on the
circumstances of detailed physical applications are conceptually
straightforward, even though specific application to the moduli phase 
spaces or holographic entanglement of black holes is reserved for a future, 
less general, report.

In passing, and because it fits naturally with the computational technique 
involved, the corresponding quantum R\'{e}nyi entropies \cite{wlodarz} are 
also evaluated explicitly here for the same prototype system, to illustrate 
the broad fact that these entropies are majorized by the Gibbs-Boltzmann 
entropy, and thus also by the bound discussed here. 
R\'{e}nyi generalized entropies were originally introduced as a measure of 
complexity in optimal coding theory \cite{wlodarz}, and have been applied 
to turbulence, chaos and fractal systems, as well as semi-inclusive 
multiparticle production \cite{jizba,varga}; however, apparently, 
they have not attained significance in black hole physics yet, 
nor in current noncommutative geometry efforts.

\section{Shannon and Boltzmann-Gibbs entropy in phase space} 
For a continuous distribution function $f(x,p)$ in phase space,
the classical (Shannon information) entropy is
\be
S_{cl}=-\int dx dp ~  f ~ \ln (f) . \label{shannonE}
\ee

For a given distribution function $f(x,p)$, 
without loss of generality centered at the origin, normalized,
$\int dx dp f=1$, and with a given variance, 
$\sigma^2= \langle x^2+p^2 \rangle =\int dx dp (x^2+p^2)f $, it is 
evident from 
elementary constrained variation of this $~S_{cl} [f]$ w.r.t.\ $f$, 
\cite{ash} (also see \cite{rajagopal}), 
 that it is maximized by the Gaussian,  
$f_g= \exp (-(x^2+p^2)/\sigma^2) / \sigma^2 \pi$, ~ 
to $S_{cl} = 1+ \ln ( \pi \sigma^2 )$. 

That is, a Gaussian represents maximal disorder and minimal information---
in thermodynamics, least dispersal energy would be available. 

Thus, it leads to a standard result in information theory \cite{ash}, 
Shannon's inequality, 
\be 
S_{cl}\leq \ln ( \pi e \sigma^2 ) ~,
\ee
which provides an upper bound on the lack of information in such distributions.

Note that, in general,  $S_{cl}$ is unbounded above,
as it diverges for delocalized distributions, $\sigma \rightarrow \infty$,
containing no information. In contrast to the Boltzmann-Gibbs entropy, it is also
unbounded below, given ultralocalized peaked distributions ($\sigma 
\rightarrow 0$), which reflect complete order and information. 

\vskip 0.5cm

In quantum mechanics, the sum over all states is given by the standard 
von Neumann entropy \cite{fano}  for a density matrix $\rho$,
\be
0\leq S_q=- \hbox{Tr} ~\rho ~ \ln \rho =-\langle \ln \rho \rangle  ~. 
\ee
This transcribes in phase space \cite{groenewold,braunss} through
the Wigner transition map \cite{book} to 
\be
0\leq S_q=-  \int dx dp ~f ~\ln_{\*} (hf)~, \label{SQ}
\ee
where the $\star$-product \cite{groenewold} 
\begin{equation}
\star \equiv e^{{\frac{i\hbar }{2}}(\stackrel{\leftarrow }{\partial }_{x} 
\stackrel{\rightarrow }{\partial }_{p}-\stackrel{\leftarrow }{\partial }_{p} 
\stackrel{\rightarrow }{\partial }_{x})}\;,
\end{equation}
serves to define $\star$-functions, such as the $\star$-logarithm, above,
e.g.\ through $\star$-power expansions, 
\be
\ln_{\*} (hf)\equiv -\sum_{n=1}^\infty  {(1-h f )^n_\*    \over n}  ~.  
\label{expansion}
\ee

Braunss \cite{braunss} has argued that, for $S_{cl}$ defined by 
$S_q +\ln h$ in the limit that the Planck constant $\hbar \rightarrow 0$, 
\be
0\leq S_q \leq S_{cl} -\ln h ~.
\ee

The logarithmic offset term relying on the Planck constant $h$ accounts 
for the scale \cite{wehrl} of the phase-space area element $dx dp$ 
in (\ref{SQ}). This scale, $h$, should 
divide $dx dp$ to yield a dimensionless phase space area element; 
Correspondingly, it should then multiply $f$, to preserve `probability', 
$\int dx dp f=1$, in the Wigner transition map from the density matrix 
$\rho$ to the Wigner Function $f$. 
E.g., for a pure state \cite{book},
\begin{equation}  
f(x,p)={\frac{1}{ h }}\int \!dy~\psi^{*}\left (x-{\frac{1}{2}} y 
\right )~e^{-iyp/\hbar } ~\psi \left (x+{\frac{1}{2}}y\right )\;.
\end{equation} 
The classical limit normally entails variations of phase-space variables 
on scales much larger than $\hbar$. Therefore these variables are normally 
scaled down to scales matched to such activity. As illustrated explicitly
in the next section, comparing quantum and classical entropies relies on 
the above offset.    
The upper bound in this Braunss inequality reflects the loss of quantum 
information involved in the smearing implicit in the classical 
limit\footnote{
Readers unfamiliar with the classical limit might find
loss of the quantum uncertainty of the theory counterintuitive and discordant 
with the loss of information involved. Actually, {\em the resolution 
to access the uncertainty} is sacrificed in this limit.
A standard consequence of the Cauchy-Schwarz 
inequality for Wigner functions is $|f|\leq 2/h$, \cite{book},   
reflecting the uncertainty principle: the impossibility of localizing $f$  
in phase space, through a delta function. The best one can do is to  
take a pillbox cylinder of base $h/2$ and height $2/h$, properly
normalized to $1=\int dx dp f$. Now, scaling   the phase-space variables 
down and $f$ up (to preserve this normalization---the volume of the pillbox,
as in the above discussion of the offset)  
ultimately collapses the base of the pillbox to a mere point in phase space; 
and leads to a divergent height for $f$, a delta function, 
characteristic of a perfectly localized classical particle. 
However, several different quantum configurations will reduce to this 
same limit: it is this extra quantum information on $h$-dependent features,
e.g.\ interference,  that is obliterated in the limit.}, effectively regarded 
as an extreme limit of subadditivity \cite{wehrl}. 

Combined with Shannon's bound, this now amounts to 
\be    
\fbox{ $ 
0\leq S_q \leq \ln \left ( {e \sigma^2  \over 2\hbar} \right ) 
$ }~,  \label{new}
\ee
i.e., the entropy is bounded above by an expression involving the 
variance of the corresponding classical limit distribution function.
It readily generalizes to multidimensional phase space ($R^{2N}$, in which 
case the logarithm is evidently multiplied by $N$, in evocation of 
Bekenstein's bound), and contexts where more information (e.g., on asymmetric 
variances) happens to be available, or refinement desired.

By virtue of (\ref{expansion}), the quantum entropy is recognized as
an expansion 
\be
S_q= \sum_{n=1}^\infty  {\langle (1-\rho)^n \rangle \over n}= 
\sum_{n=1}^\infty  {\langle (1-h f )^n_\* \rangle \over n}.  
\ee
The leading term, $n=1$, ~$1-$Tr$\rho^2=\langle 1-hf\rangle$, 
is the {\em impurity} \cite{groenewold,fano,book},
often referred to as linear entropy.
Like the entropy itself, it vanishes for a pure state 
\cite{groenewold,fano,book},
for which $\rho^2=\rho$, or, equivalently, $f\* f = f/h$. Each term in the 
above expansion
then projects out $\rho$, or $\* hf$, respectively: {\em pure states saturate 
the lower bound on} $S_q$.

A likewise additive (extensive) generalization of the quantum 
entropy is the R\'{e}nyi entropy \cite{wlodarz},    
\be 
R_\alpha = \frac{1}{1-\alpha} \ln  \langle  \rho^{\alpha-1} \rangle 
=  \frac{1}{1-\alpha}
\ln  \int \frac{dx dp}{h}  (h f )^\alpha_\*  ~,
\ee
where the limit $\alpha\rightarrow 1$ yields $R_1=S_q$, and the above-mentioned 
impurity is $ 1-\exp (- R_2)$. For continuous distributions (infinity of 
components) discussed here, $R_0$ is divergent. 

For $\alpha\geq 1$, $R_\alpha\geq R_{\alpha+1}$, so $S_q \geq R_{\alpha}$, and 
it is also bounded below by 0  \cite{wlodarz}, i.e., 
\be
S_q \geq R_{\alpha} \geq R_{\alpha+1}\geq 0 ~,  \label{renyitrain} 
\ee
so that, a fortiori, the R\'{e}nyi entropy is also bounded by (\ref{new}).

\section{Gaussian Illustration}
To illustrate the above inequalities,  
consider the Gaussian Wigner Function of {\em arbitrary} half-variance $E$,
\be
f(x,p,E) =\frac{e^{-\frac{x^2+p^2}{2E}}}{2\pi E }
= e^{-\frac{x^2+p^2}{2E}-\ln (2\pi E)}.
\label{wiggie}
\ee
This happens to be the phase-space Wigner transform of a Maxwell-Boltzmann 
thermal distribution for a harmonic oscillator \cite{bartlett}, in 
suitably rescaled units, normalized 
properly to unity, and with mean energy $E=\langle (x^2+p^2)/ 2 \rangle$. 

Calculation of the entropy of this distribution,  is, of course, a freshman 
physics problem, but its independent phase-space 
derivation  \cite{wang} (also see  \cite{agarwal}), is reviewed here, i.e., 
evaluation of (\ref{SQ}) directly.

For $E=\hbar/2$, the distribution reduces to just $f_0$, the Wigner 
Function for a 
pure state (the ground state of the harmonic oscillator). Hence 
\cite{groenewold,book},
\be
f_0\* f_0=  {f_0\over h }~, 
\ee
so that $f_0$ is $\*$-orthogonal to each of the terms in the sum 
(\ref{expansion}),
and hence $S_q=0$, indicating saturation of the maximum possible 
information content.

For generic width $E$, the Wigner Function $f$ is not that of a pure state, 
but it still happens to always amount to a $\star$-exponential \cite{imre} ~ 
($e_\*^a\equiv 1+ a + a\* a /2!+ a \* a \* a/3!+...$)  as well, 
\begin{equation} 
hf= e^{-\frac{x^2+p^2}{2E}+\ln (\hbar/ E)}=
e_\* ^{-\frac{\beta}{2\hbar} (x^2+p^2)   +\ln (\frac{\hbar}{E} \cosh 
(\beta/\hbar)) },   \label{felicity}
\end{equation}
where an ``inverse temperature" variable $\beta(E,\hbar)$ is useful to define, 
\be
\tanh (\beta/2 )\equiv {\hbar\over 2E}  \leq 1  \qquad \Longrightarrow 
\qquad  \beta= \ln \frac{E+\hbar/2 }{E-\hbar/2 } ~.
\ee
(Thus the above pure state $f_0$ corresponds to zero temperature, 
$\beta=\infty$.)

Since $\star$-functions, by virtue of their $\star$-expansions,  
 obey the same functional relations as their non-$\star$ analogs,
inverting the $\star$-exponential through the $\star$-logarithm 
and integrating (\ref{SQ}) yields directly the standard thermal physics 
result,
\be
S_q( E,\hbar) = \frac{E}{\hbar}\ln \left( \frac{2E+\hbar }{2E-\hbar }\right )
 + \frac{1}{2}\ln \left (  (\frac{E}{\hbar})^2-  \frac{1}{4} \right )= \frac{\beta}{2} 
\coth (\beta/2 ) - \ln (2 \sinh(\beta/2 ) ). \label{gaussent}
\ee
Indeed, this can be seen to be a monotonically nondecreasing function of $E$,
attaining the lower bound 0 for the pure state $E\rightarrow \hbar/2$ 
($\beta\rightarrow \infty$, zero temperature).

The classical limit, $\hbar\rightarrow 0$ ($\beta \rightarrow 0$, infinite
temperature) thus follows,
\be
S_q \rightarrow 1+\ln (E/\hbar) = \ln (\pi e 2E ) - \ln 
h= S_{cl}(E)  -\ln h ~,  \label{cllim}
\ee
and is explicitly seen to bound the expression (\ref{gaussent}) 
for all $E$, saturating it for large $E>>\hbar$, in accordance with 
Braunss' bound. That is, the upper bound (\ref{new}) is saturated 
for Gaussian quantum Wigner functions with $\sigma^2 >>\hbar$.

Note the region $E<\hbar/2$, corresponding to ultralocalized spikes excluded 
by the uncertainty principle,  was not allowed by the above derivation method,
since, in this region, no $\*$-Gaussian can be found to represent the Gaussian.
(It would amount to complex $\beta$ and $S_q$, linked to 
thermal expectations of the oscillator parity operator.)

NB. An alternate heuristic proposal of ref \cite{wehrl} for the classical 
limit of the entropy effectively starts from the Husimi 
phase-space representation \cite{book}; it first effectively drops 
all $\*_H$s in (\ref{SQ}) and easily evaluates (\ref{shannonE}) instead 
(which is well-defined because $f_H\geq 0$ automatically), {\em before} 
completing the   
transition to the classical limit $\hbar \rightarrow 0$. It also, ultimately, 
yields the same answer (\ref{cllim}), since the Husimi representation 
of the Gaussian Wigner Function (\ref{wiggie}), 
\be
f_H\equiv \int dx' dp' ~\frac{e^{-({(x'-x)^2+(p'-p)^2})/\hbar} }{\pi\hbar}~ 
f(x',p') 
=\frac{e^{-\frac{x^2+p^2}{2E+\hbar }}}{\pi( 2E +\hbar) } ~, 
\ee
is also a Gaussian.  Utilized to evaluate (\ref{shannonE}), it 
yields $\ln (\pi e (2E+\hbar) )$,
which has the more direct expression 
$S_{cl}$ of (\ref{cllim}) as its classical limit. (For the ground state,
$E= \hbar/2$, which is a coherent state, this semiclassical entropy
reduces to a characteristic minimal value, $1+\ln h$.) 
 
By virtue of (\ref{felicity}), $\*$-powers of the Gaussian are also
straightforward to take, and thus the R\'{e}nyi entropies can be readily 
computed:  
\be
R_\alpha = \frac{1}{1-\alpha} \ln  \left ( \frac{(2 \sinh (\beta/2) ) ^\alpha }
{2 \sinh (\alpha\beta/2) ) } \right )
=\frac{1}{\alpha-1} \ln  \left (  \left (\frac{E}{\hbar}+\frac{1}{2}\right 
)^\alpha - \left (\frac{E}{\hbar}       -\frac{1}{2}\right )^\alpha \right ).
\ee

Note $\alpha\rightarrow 1$ checks with the above (\ref{gaussent}), 
$R_1 \rightarrow S_q$. Also, in the pure state limit, $E=\hbar/2$, it is
evident that $R_\alpha=0$ checks for all $\alpha\geq 1$.  
(For $\alpha>1$ and  the small 
disallowed values $E<\hbar/2$, $R_\alpha<0$.) 

$R_\alpha$ is also a nondecreasing 
function of $E$; and, in comportance with (\ref{renyitrain}), 
 a nonincreasing function of $\alpha$. 
Up to an additive, $\alpha$-dependent constant, the classical limit is
identical to that for the entropy itself,
\be
R_\alpha \rightarrow {\ln  \alpha\over \alpha-1} +\ln (E/\hbar)  ~, 
\ee
in agreement with the classical result of \cite{varga}. It may well be that,
as in the contexts touched upon in the introduction, specific $\alpha$s may 
well provide more detailed or practical measures of complexity in Hawking 
radiation with sparse information available.

{\em If a specific quantum Hamiltonian were actually 
available} for the system in question (a rare occurrence), then the classical
limit of the entropy of the system would be straightforward---and thus the
inequality discussed here would not be that powerful, since the classical
entropy itself would be at hand, in general lower than the Shannon bound.

For such a simple system, the upper-bounding classical entropy 
would result out of the phase-space partition function specified 
by the corresponding classical hamiltonian (the Weyl symbol of the 
quantum hamiltonian).  This is easily illustrated explicitly by 
hamiltonians which are positive $N$-th powers of the oscillator 
hamiltonian, so that, simply,  
\be
f_{cl} \propto \exp ( -( (x^2+ p^2) /2E)^N).
\ee 
The bounding classical entropy then reduces by standard thermodynamic 
evaluation to be just (\ref{shannonE}), 
\be
 S_{cl}= \frac{ 1}{N}  + \ln \left ( 2 \pi E~ \Gamma \left( 1+ \frac{1}{N}
\right ) \right ),
\ee
lower than the cooresponding Shannon bound, 
\be
1+\ln \left ( \pi E  \frac{\Gamma (1+ 2/N)}{\Gamma (1+ 1/N)}\right ) ~.
\ee

{\em Note Added~~} A referee has identified a technical gap in 
Braunss' formal proof of his inequality in \cite{braunss}, 
which is, nevertheless, assumed here. 

\vskip 0.3cm

\noindent{\it This work was supported by the US Department of Energy, 
Division of High Energy Physics, Contract DE-AC02-06CH11357;  and the 
Collaborative Project GEP1-3327-TB-03 of the US Civilian Research
and Development Foundation. 
Helpful discussions with A Polychronakos, T Curtright, 
and G Jorjadze are acknowledged.}

\end{document}